\documentclass[prk,showkeywords,twocolumn]{revtex4-1}
\usepackage{graphicx}
\usepackage{epsfig}
\usepackage{psfrag}

\usepackage{hyperref}
\hypersetup{colorlinks=true,linktoc=page,linkcolor=blue,citecolor=red,urlcolor=cyan}

\newcommand{\neff}{N_{\text{eff}}}
\newcommand{\higgs}{\langle v \rangle}

\begin{document}

\title{The cosmological lithium problem, varying constants and the $H_0$ tension}
\author{S. A. Franchino-Vi\~nas$^{1,2}$}
\author{M. E. Mosquera$^{2,\,3}$}
\affiliation{$^1$\small\it Institut f\"ur Theoretische Physik, Universität Heidelberg, D-69120 Heidelberg, Germany.}
\affiliation{$^2$\small\it Departamento de Física, Fac. de Cs. Exactas, Universidad Nacional de La Plata, c.c.~67 1900, La Plata, Argentina.}
\affiliation{$^3$\small\it Facultad de Ciencias Astron\'omicas y Geof\'{\i}sicas, Universidad Nacional de La Plata. Paseo del Bosque S/N 1900, La Plata, Argentina.}

\begin{abstract}
In this work we show that the cosmological lithium problem and the $H_0$ tension could be eased at the same time by allowing variations in the fundamental constants. We compute the primordial abundances of light elements resulting from Big Bang Nucleosynthesis considering the fine structure constant, the Higgs' vacuum expectation value and Newton's constant as free parameters. Using the observational data for abundances, we set constraints on the variations of the fundamental constants. An interpretation of the  results in terms of the number of effective relativistic species gives $N_{\rm eff}=4.04\pm0.12$. If one extrapolates the fit of {\sffamily Planck} considering this $\neff$, the value of the inferred Hubble's constant shifts to $H_0=(71.85\pm0.77)\,\text{km}\,\text{s}^{-1} \text{Mpc}^{-1}$, compatible with current direct determinations.
\end{abstract}


\maketitle

\section{Introduction}

The Big Bang Nucleosynthesis (BBN) process that yields the primordial abundances of light elements is by now well known \cite{Cyburt:2015mya}. The standard BBN depends only on one free parameter, the baryon-to-photon ratio $\eta_B$. Under usual assumptions $\eta_B$ remains the same until the recombination epoch, so its value can be inferred from the Cosmic Microwave Background (CMB) data \cite{wmap13,planck18}. Afterwards, a numerical computation involving almost one hundred coupled differential equations  provides remarkable precise predictions for the primordial abundances of $^4$He and D. 

However, its original success further relied (among others) on the following two facts:
\begin{itemize}
 \item the assumption that the so-called fundamental constants have the same value today and at the time of BBN, i.e. at both extremes of the universe;
 \item the belief that the discrepancy by a factor two (or even larger) between predictions of the primordial lithium abundance and the corresponding observational data can be disregarded.
\end{itemize}
In this work, we will instead adopt a more recent posture, which considers that the latter has become a major issue that must be solved: the so-called cosmological lithium problem \cite{Fields:2011zzb}. Our goal is  to show that we can provide a solution to it by relaxing the first assumption stated above and using the observational data as guidance. As a by-product, this will alleviate the increased tension that has been generated in the last years between direct and indirect determinations of Hubble's constant $H_0$ \cite{DiValentino:2020zio, Verde:2019ivm}.

Our motivations are the following. Considering a variation of fundamental constants can be seen as an agnostic way to analyze different possible scenarios at once, since it does not rely on a specific model. Such variations could be \emph{a posteriori} ascribed, for example, to some systematics, uncertainties, running of couplings in the standard model of particles or simply some new physics. Although for this very reason it could be criticized as too weak, we think that it can provide precious help for model building. In this scenario, different possibilities have been considered in the past \cite{uzan11}, some of them with partial success concerning the Li problem \cite{Berengut:2009js,Coc:2012xk,Martins:2020syb}. 


Regarding the latter, some original criticisms were directed to  depletion processes of lithium which, allegedly, were not correctly understood \cite{Korn:2006tv}. However, in the last decade new available measurements at low metallicity resulted in abundances below the Spite plateau \cite{hosford10,sbordone10}, thus increasing the previous discrepancy. This new data has given place to many proposals. Particle models have been built to reduce the amount of $^7$Li, mainly through Be destruction: a new particle that could either inject radiation through its decay \cite{Poulin:2015woa, Alcaniz:2019kah} or prompt nuclear reactions \cite{Goudelis:2015wpa} seems among the most promising scenarios. However, the debate is far from being settled.

\section{Formalism}

Taking the abovementioned facts into account,  we have chosen to modify the AlterBBN numerical code \cite{arbey12}, what allows us to compute the primordial abundances while considering possible variations of a set of fundamental constants detailed below. We have computed $\eta_B$ as customarily done, considering the CMB temperature 
from \cite{Fixsen:2009ug} and a fixed baryon density, which we take as the weighted mean value of the base-$\Lambda$CDM results in {\sffamily WMAP} \cite{wmap13} and {\sffamily Planck} \cite{planck18}. We have thus kept the assumption that $\eta_B$ does not evolve between BBN and CMB epochs. Then, we have inferred the values of the fundamental constants at BBN performing a $\chi^2$ statistical test: we have required the numerical yields to provide the best possible fit to the observational data, including that of lithium on an equal foot. We assume that the fundamental constants do not evolve inside BBN, what seems reasonable given the narrow period in which it takes place; at the same time, this allows for largely model-independent results.

More technically, the code was modified in order to include a possible variation of the gravitational constant $G_N$, by including a free parameter ($G_{\rm BBN}$) in Friedman's equation. In order to make this variation meaningful, we have kept the QCD mass scale ($\Lambda_{\rm QCD}$) fixed. Notice that once we neglect dark and nonrelativistic sources in Einstein's equation, the modification of $G_N$ is tantamount to considering a variation in the number of effective relativistic species at BBN  ($\neff$) \cite{Barger:2003zg}. An increase in the gravitational constant implies a faster expansion rate, which is crucial in the determination of the reactions' freezeout. This has been widely studied, e.g., for the neutron-to-proton conversion \cite{bernstein89,esma91}; around the standard BBN, it results in smaller (larger) yields for $^7$Li (D and $^4$He). 

Additional refinements to study the variation of the fine structure constant ($\alpha$) and Higgs vacuum expectation value ($\higgs$) were introduced as explained in Refs. \cite{landau08,nollet02,iguri99,landau06,mosquera17,scoccola07,civitarese10} and references therein. In particular, we have considered the dependence of the deuterium binding energy on $\higgs$ derived in Ref. \cite{civitarese10}, and employed the value obtained by using the Argonne $v_{18}$ potential. 

On the observational side, to set constraints on the variations we have used the abundances of deuterium \cite{zavarygin18,cooke18,riemer17,cooke16,balashev16,cooke14,pettini12,noterdaeme12,balashev10,ivanchik10}, helium \cite{valerdi19,kurichin19,fernandez18,cooke18b,izotov14,izotov13,aver13,aver10,izotov10} and lithium \cite{gruyters16,monaco12,mucciarelli12,nordlander12,nissen12,mucciarelli11,melendez10,sbordone10,hosford10}, including also the already mentioned measurements below Spite's plateau.  To avoid possible systematics, the only criterion that we have followed in choosing the data regards their date of publication: all of them correspond to the last decade. We have checked the  consistency of the data \cite{pdg} and found that for helium and lithium one must respectively increase the observational errors by a factor $\Theta_{\rm He}=1.82$ and $\Theta_{\rm  Li}=1.93$.

\section{Results and discussion}

\begin{table*}[t]
\renewcommand{\arraystretch}{1.4}
\setlength\tabcolsep{10.8 pt}
\begin{center}
\caption{Best-fit of the fundamental constants. $N$ is the total number of observational data and $x$ stands for the number of free parameters. {The uncertainties correspond to a 68\% confidence level.}} \label{tabla:ajuste}
\begin{tabular}{c c c c c } \hline\hline
&$\Delta \alpha/ \alpha [10^{-2}]$ &$\Delta \higgs/ \higgs [10^{-2}]$ & $G_{\rm BBN}/G_N$ &  $\chi^2_{min}/(N-x)$  \\ \hline 
Fit 1 &$-2.83 \pm 0.24$ & $--$ & $1.167^{+0.015}_{-0.016}$ & $1.95$ \\ 
Fit 2 &$-3.35 \pm 0.45$ & $-1.50 \pm 0.20$ & $1.170 \pm 0.020$ & $0.70$  \\ \hline\hline
\end{tabular}
\end{center}
\end{table*}

\begin{table*}[t]
\renewcommand{\arraystretch}{1.4}
\setlength\tabcolsep{10.8 pt}

\begin{center}
\caption{Primordial $^4$He mass fraction ($Y_{\rm P}$) and primordial abundances of D and $^7$Li relative to hydrogen (D/H and $^7$Li/H respectively). The following cases are included: the weighted mean value of the considered observational data, the standard computation and the prediction using variations of fundamental constants. The uncertainties for the observational values are increased as explained in the text. {The uncertainties of the numerical yields correspond to a 68\% confidence level.} } \label{tabla:abundancias}
\begin{tabular}{c c c c } \hline\hline
& ${\rm D/H}\, [10^{-5}]$ & $Y_{\rm  P}$ & ${^7\rm Li/H}\, [10^{-10}]$ \\ \hline 
Fit 1 & $2.605 \pm 0.040$ & $0.2472 \pm 0.0016$ & $2.09 \pm 0.12$ \\
Fit 2 & $2.545 \pm 0.058$ & $0.2561 \pm 0.0050$ & $2.03 \pm 0.22$ \\ \hline
Observational weighted mean value& $2.541 \pm 0.022$ & $0.2542 \pm 0.0014$ & $2.01 \pm 0.14$\\ \hline
Standard computation& $2.547 \pm 0.050$ & $0.2466 \pm 0.0001$ & $4.60 \pm 0.32$\\ \hline\hline
\end{tabular}
\end{center}
\end{table*}

The results of the $\chi^2$-test, i.e. the most probable values of the fundamental constants at BBN according to our statistical analysis, are shown in Table \ref{tabla:ajuste}.
Notice that the variation of a fundamental constant $X$ is casted in terms of $\Delta X=X_{\rm BBN}-X$, i.e. subtracting the reference value in Ref. \cite{pdg} to the one obtained in the fit. 

In Table \ref{tabla:abundancias} we present the predicted abundances for our fits, together with the results of standard BBN {(computed with AlterBBN)} and the mean weighted value of the used observational data. {Notice that as a result of our weighting, the observational yields are slightly more stringent than in usual analysis. }

The joint variation of $\alpha$ and $G_N$ already implies a significant improvement in the fit (cf. the Fit 1 results in Table \ref{tabla:abundancias}), which can not be obtained by merely varying one parameter at a time. In this case the best-fit primordial abundances of $^7$Li and D are in agreement with the observational data, however the $^4$He abundance shows a mild discrepancy with the observational data at the $2\sigma$ level. 
In any case, our fit provides a slightly better agreement with observations than the standard computation.

The scenario is further improved once the Higgs' vacuum expectation value is allowed to vary: in our best fit the primordial abundances of all the light nuclei show a perfect agreement with the observational data (cf. the Fit 2 results in Table \ref{tabla:abundancias}). In addition, it is interesting to notice that the corresponding $G_{\rm BBN}$ is to a great extent independent on the inclusion of a free $\higgs$.
Now the question is if such modifications in the fundamental constants can be justified. 

\begin{figure*}[t]
 \begin{center}
  \includegraphics[width=1.0\textwidth,height=0.7\textwidth]{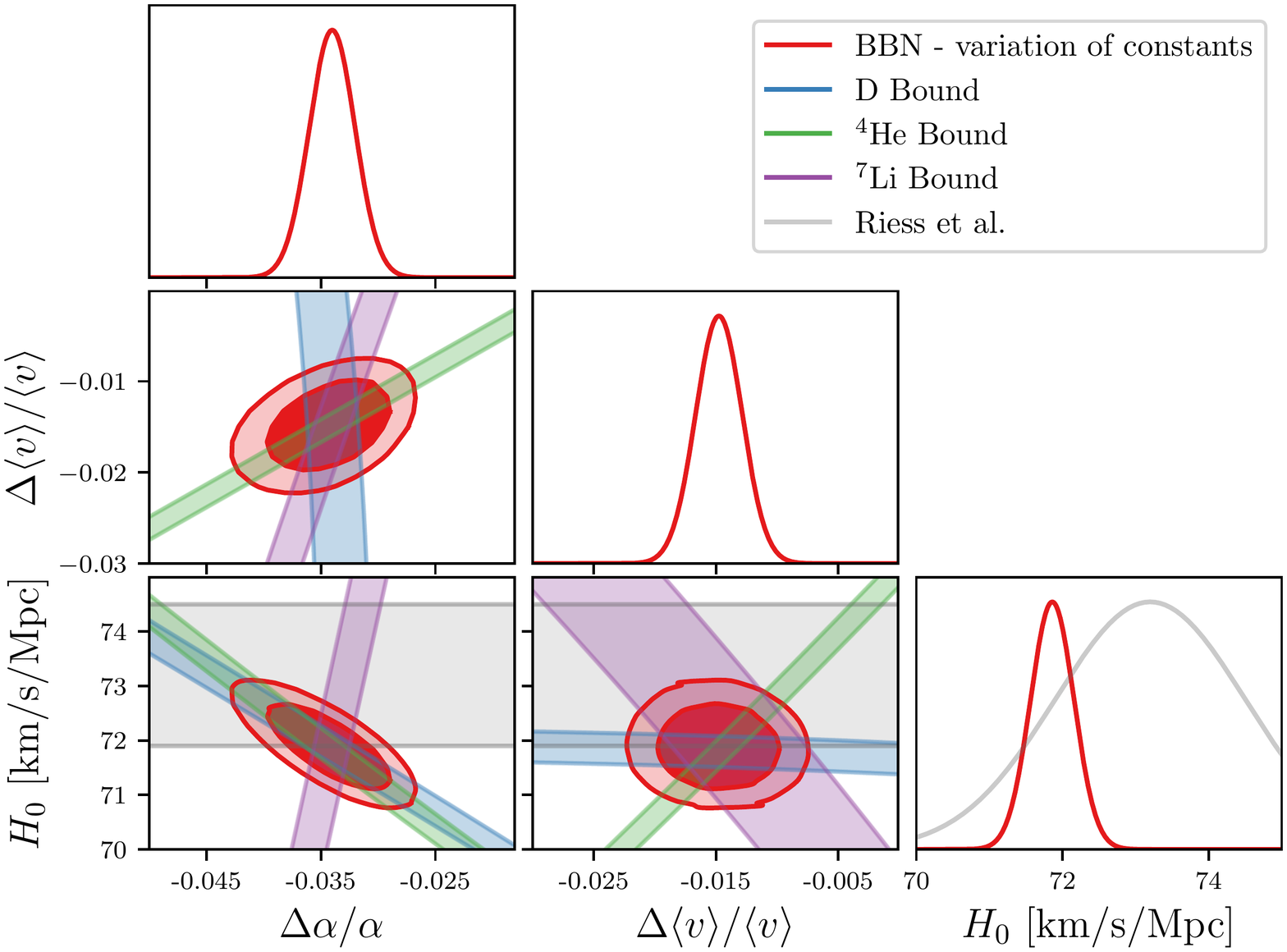}
 \end{center}
 \caption{Contour plots and likelihoods. 
 The red contours correspond to our statistical analysis of the Fit 2 (second entry in Tables \ref{tabla:ajuste} and \ref{tabla:abundancias}) at 68\% and 95\% confidence level. The grey regions are given by the $H_0$ obtained in Ref. \cite{Riess:2020fzl} considering Cepheids (at 68\% confidence level). The blue, green and violet zones are the observational $1\sigma$-level constraints of deuterium, helium and lithium respectively.}
 \label{fig:alpha_vs_H}
\end{figure*}

\subsection{Running coupling constants}
One interesting candidate is the running that coupling constants develop in quantum field theories when one considers the quantum corrections of interacting particles at an energy different from the reference one. 

Notice first of all that runnings are not just a theoretical tool: as an example, the running of the fine structure constant has been recently confirmed by the experimental data from the {\sffamily KLOE} detector in the region around $1$ GeV \cite{KLOE-2:2016mgi}. Running is predicted to take place at BBN, since energies in the relevant processes can be taken to be of the order of BBN temperatures, i.e. $T\sim0.03-1\, \text{MeV}$. All the more, there could {exist an enhancement of this effect} as a consequence of the large tail of available highly-energetic photons.  Another major aspect is that running would have no effect at recombination time, where possible fine structure variations are already constrained below the percent level using {\sffamily Planck} and {\sffamily Planck} {in combination with baryonic acoustic oscillations (BAO)} data \cite{Hart:2019dxi}. 

The disadvantage is that runnings are expected to be process-dependent once one consider contributions outside from the universal log terms \cite{Donoghue:2019clr}. For example, in the realm of the Standard Model of particle physics the fine structure constant shows different behaviours for space- and time-like processes \cite{Jegerlehner:2011mw}. The situation worsens once we recognize that we are considering one single variation of $\alpha$, propagating through all the chain of nuclear reactions, what clearly is an oversimplification. An analogous situation takes place when implementing the variation of $\higgs$: in it we are encoding several different runnings that otherwise should involve precise computations which are not available. As a consequence, our discussion of runnings should be seen as a feasibility study rather than a precise determination. After this disclaim, it is appealing to observe that the theoretical computations, due to quantum contributions of electrons and positrons, involve a resonance in the running of $\alpha$ at BBN temperatures; a rough estimation \footnote{In such affirmation we are assuming that space-like processes do not take place much more frequently than time-like processes at BBN.}  tells us that it has the correct sign but would account only for a 10\% of the observed $\Delta \alpha$ \cite{Jegerlehner:2011mw}. In order to carry out a more detailed analysis, we should wait until more precise {experimental/numerical} data becomes available.

In what respects $G_N$, its running has been recently theoretically analyzed in several models of quantum fields in curved spaces \cite{Bosma:2019aiu, Franchino-Vinas:2018gzr, Bjerrum-Bohr:2016hpa}. A modification on the percent level is precluded in these scenarios by the smallness of Planck's mass. In order  to reproduce such large variations one has to resort to modified models of gravity of diverse level of complexity, like scalar-tensor theories \cite{Ballesteros:2020sik, Brans:1961sx} or the running vacuum models \cite{Sola:2016jky, Sola:2017znb}. 
More appealing to us is its interpretation in terms of $\neff$.

\subsection{Varying $G_N$ vs. $\neff$}
 Indeed, the possibility to include additional light degrees of freedom has been widely considered in the literature \cite{DiValentino:2021izs}, given that it provides a way {to increase $H_0$} and decrease the sound horizon, while leaving unaffected their product, thus partially avoiding discrepancies with the strong constraints placed by BAO observables. 
Let us sketch this possibility in a simplified computation that will be enough for our purposes.

If one neglects all but the radiation contribution in Friedmann's equation, one obtains the equivalence \footnote{One can define $\Delta \neff=\neff-\overline N$, being either $\overline N= N_{\nu}=3$ the number of neutrino's generations in the standard model of particles, or $\overline N=3.046$, the value of the cosmological standard model. Our subsequent discussion is not affected by such a choice.} 
\begin{equation}\label{eq:neff}
\frac{\Delta G}{G_N}=\frac{7}{43} \Delta \neff \, .
\end{equation}
Analogously, from Friedmann's equation one can compute the change in $H_0$ induced by an increase in $\neff$. As a way to avoid two of the most stringent cosmological constraints, we fix the radiation density {to a given value}\footnote{The exact value is not important, since it cancels in the computation.} and require the product $r_{s}H_0$ of the sound horizon at drag epoch times Hubble's constant to be unaffected.  Following the steps of Ref. \cite{carneiro18} we obtain thus Hubble's constant as a function of $\neff$:
\begin{eqnarray}\label{eq:H0}
 H_0(\neff)\approx H_0(3)\sqrt{0.595+0.135\neff}\,.
\end{eqnarray}

Turning back to our fits, both correspond roughly to $\Delta \neff \sim 1$, i.e. to the inclusion of an effective additional relativistic degree of freedom. Such a variation, being compatible with the results available from {\sffamily WMAP} \cite{wmap13} and their combination with those of the South Pole Telescope \cite{Hou:2011ec}, is disfavoured at least at the $3\sigma$ level by the study of {\sffamily Planck}, for which $\neff^{\text{{\sffamily Planck} TT+lowE}}\sim3.00^{+0.57}_{-0.53}$ \cite{planck18} at a 95\% confidence level. 

A careful analysis unveils that the reason behind this discrepancy among distinct CMB determinations is the high resolution that {\sffamily Planck} has achieved in the tail of the CMB spectra \cite{Knox:2019rjx,Follin:2015hya,Baumann:2015rya}. This has been shown to be  also responsible for the existence of a mild internal tension of $2\sigma$ in the inference of matter density from low and high multipole momenta, generating some debate \cite{Knox:2019rjx, Addison:2015wyg}. Outside from this comment, there are arguments to tilt the balance in both directions. 

On the one hand, a fit of the CMB data with $\neff$ fixed to such values would increase the tension in other sectors. As an example, it is expected to affect {the fit of } the matter power spectrum, increasing the value of $\sigma_8$ (the RMS mass fluctuation on scales $8h^{-1}\text{Mpc}$, $H_0=:100 h\,\text{km}\,\text{s}^{-1} \text{Mpc}^{-1}$)
, thus tensing the already tight situation. However one could in principle try to circumvent these constraints by turning on interactions and permitting variations of other cosmological parameters \cite{Kreisch:2019yzn,Escudero:2019gvw}.

On the other hand, under certain additional assumptions our result is compatible with the existence of a sterile neutrino. This idea has been mainly motivated by the neutrino anomalies first detected in {\sffamily LSND} \cite{Athanassopoulos:1997pv} and confirmed by {\sffamily MiniBooNE} \cite{Aguilar-Arevalo:2018gpe}. 
An additional key point is that, {keeping the other cosmological parameters fixed,} $\Delta \neff \sim 1$ is the right amount of change we need in order to solve the $H_0$ tension \cite{DiValentino:2021izs, planck18}. 

Indeed, with the aid of eqs. (\ref{eq:neff}) and (\ref{eq:H0})  we can recast our results in terms of $H_0$. Considering one of the worst possible scenarios, i.e. fixing $H_0(3)$ to the baseline $\Lambda$CDM result from {\sffamily Planck} \cite{planck18},
our Fit 2 gives {at the 68\% confidence level}
\begin{equation}
H_0^{\Delta \neff}=(71.85\pm0.77)\,\text{km}\,\text{s}^{-1} \text{Mpc}^{-1}.  
\end{equation}
This is in appreciable agreement with the value $H_0^{\text{Ceph.}}=(73.2\pm1.3)\,\text{km}\,\text{s}^{-1} \text{Mpc}^{-1}$ (68\% confidence level), which has been recently obtained using Cepheids' data \cite{Riess:2020fzl}.

We provide more detailed information regarding the Fit 2 in Fig. \ref{fig:alpha_vs_H}, where the corresponding constraints in all the different planes defined by $\Delta\alpha$, $H_0$ and $\higgs$ are displayed \footnote{The likelihood of $H_0$ corresponding to Ref. \cite{Riess:2020fzl} has been modelled as a Gaussian.}. We also show the single element bounds derived from the observational data at the $1\sigma$ level as well (we are fixing the non-plotted variable to its Fit 2 value). 
In every plane all abundances intersect in a non-empty region {which, when suitable,} has a substantial overlap with the Cepheids result at the $1\sigma$ level. 

\section{Final remarks}
We have obtained the following nontrivial results. First, BBN \emph{per se} does not preclude a non-vanishing $\Delta \neff$: the delicate equilibrium among abundances can allow for an extra relativistic species if one considers additionally small modifications in $\alpha$ or $\higgs$. {This is valid at least if one considers the observational abundances gathered during the last decade.} In light of this fact, it will be interesting to reanalyze joint CMB+BBN fits and the conclusions drawn in scenarios where, instead of CMB, BBN is used to disentangle the $H_0$ data from BAO \cite{Schoneberg:2019wmt,Cuceu:2019for}. 

Secondly, in a largely model-independent way we have found that one can also reconcile theoretical predictions of BBN with the lithium observational data. From our perspective, the additional constraints from the lithium abundance play the role of a Hubblemeter, {given that they are almost perpendicular to those of the other elements in the $H_0$-$\alpha$ plane}.  

Our analysis requires variations of $\alpha$ and $\higgs$ on the percent level, which are feasible even if not completely understood, while it prefers $\Delta\neff\sim 1$. The latter is stable under the inclusion of $\higgs$ as a free parameter and reminds us the analysis of neutrino anomalies \cite{Aguilar-Arevalo:2018gpe}. The more stirring fact is that such $\Delta\neff$ is {the right amount one} would need to ease the $H_0$ tension from the CMB side. Our results could therefore be interpreted as further support to the existence of additional relativistic species. Of course, a complete statistical analysis of CMB+BBN should be performed to validate or refute this assertion, using for example the $\Lambda$CDM model with the appropriate variations of constants. 

\section*{Acknowledgments}
The authors greatly appreciate comments received from H. Falomir, A. Gómez-Valent, F.D. Mazzitelli and L. Verde, and the subsequent discussions. S.A.F. is grateful to G. Gori and the Institut für Theoretische Physik, Heidelberg, for their kind hospitality. M.E.M. is supported by a grant (PIP-616) of the National Research Council of Argentina (CONICET). M.E.M. is member of the Scientific Research Career of the CONICET. S.A.F. acknowledges support by UNLP, under project grant X909 and ``Subsidio a Jóvenes Investigadores 2019''.

\bibliography{gn2}
\end{document}